\documentstyle[prc,aps,epsf]{revtex}
\begin{document}
\draft


\title{Inclusive Particle Spectra at RHIC}  

\author{D.~E.~Kahana$^{2}$, S.~H.~Kahana$^{1}$}
 
\address{$^{1}$Physics Department, Brookhaven National Laboratory\\
   Upton, NY 11973, USA\\
   $^{2}$31 Pembrook Dr., Stony Brook, NY 11790, USA}
\date{\today}  
  
\maketitle  
  
\begin{abstract}
A simulation is performed of the recently reported data from PHOBOS at
energies of $\sqrt{s}=56,130 \,A$ GeV using the relativistic heavy ion cascade
LUCIFER which had previously given a good description of the NA49 inclusive
spectra at $\sqrt{s}=17.2 \,A$ GeV. The results compare well with these 
early measurements at RHIC.
\end{abstract}

\pacs{25.75, 24.10.Lx, 25.70.Pq}

The Relativistic Heavy Ion Collider (RHIC) at Brookhaven National Laboratory
was constructed with the explicit purpose of creating and analysing a form of
hadronic matter referred to as quark-gluon plasma. Certainly partons, when
struck with sufficient energy, may acquire enough momentum to travel beyond
the confines of their host hadron. In $p+p$ experiments at the RHIC energy of
$\sqrt{s}\sim 200 \,A$ GeV the contribution of such `jets' to the inclusive
production of  mesons is not large, perhaps less than $5\%$
\cite{jets}. Nevertheless, sufficient thermal energy can possibly be pumped
into a massive ion-ion system, via genertation of the less well defined
mini-jets \cite{eskola}, to free or create large numbers of partons in an
ion-ion collision. The existence and precise nature of any ensuing phase
change, from infinite hadronic to partonic matter \cite{infiniteplasma}, is
still the subject of debate. Truly macroscopic systems in which plasma might
be realised do exist in nature, in the early universe \cite{earlyuniverse} or
in a neutron star \cite{sn87a}. Although for a finite system the question
whether an actual phase change occurs may be somewhat academic, one might
still hope to identify a deconfined mode by sufficiently sharp rather than
truly discontinuous changes in appropriate observables. For example, the
transverse energy measured in an ion-ion collision can be used to define, in
a model, the system temperature and the relationship to say the density, of
the number of mid-rapidity pions as established by experiment. The hadron
number density is a measure of the entropy created in the collision, a
quantity definable even for a non-equilibrium finite system, and one
reasonably expected to be highly sensitive to the increase in degrees of
freedom accompanying parton deconfinement. The overreaching concern is, then,
how to identify a meaningful variation in dependences between relevant
observables.

Here, we address only the most recent and remarkably prompt measurements by
the PHOBOS \cite{PHOBOS} collaboration at RHIC. The highly successful, early
running of the RHIC facility, albeit at lower than the ultimate energy and
luminosity, together with this efficient small detector have already provided
the heavy ion community with interesting, perhaps even provocative
results. PHOBOS, lacking for the moment particle identification or momentum 
determination, is initially limited to a measurement of the charged particle
density in pseudo-rapidity $\frac{dN}{d\eta}$. We analyse the PHOBOS results
theoretically with the hadronic cascade LUCIFER \cite{LUCIFERI,LUCIFERII},
adopting the position that this analysis simply presents an extrapolation
from the earlier NA49 inclusive measurements \cite{NA49} to the considerably
higher energy RHIC determinations.

It seems appropriate to compare these initial observations at RHIC with
simulations which assume no plasma is present.  The purest such comparison
would employ a model involving only hadronic degrees of freedom. A recent
comparison does exist with the partonic code HIJING \cite{wang1}. The
instrument for the present exploration of the RHIC domain is the code
LUCIFER, described in detail elsewhere \cite{LUCIFERII} and available by
downloading from a BNL theory home page. Suffice it to say that this
simulation was prepared for use at relativistic energies attainable at RHIC
and tested against the CERN SPS heavy-ion experiments. This purely hadronic
simulation gave a good account of the two general particle production
experiments at the SPS, those for S+U and for Pb+Pb
\cite{LUCIFERII,NA49,NA35}. Thus LUCIFER might be used as a standard against
which to place the very interesting results from PHOBOS, a means for defining
the `ordinary' in proceeding from the SPS to RHIC.  This can be
accomplished by a slight tuning of LUCIFER multiplicities to provide very
close to quantitative agreement for the SPS $h^-$ rapidity spectrum.  In
retrospect \cite{LUCIFERII}, the predictions for the latter spectrum were
perhaps $10-15\%$ high when compared with the latest NA49 $h^-$ determination
\cite{NA49b}.
 
One possibility, exploited in our methodology, is that to some extent an
ion-ion collision is describable by multiple interactions between excited
hadrons only. In such a picture the constituent quarks are excited to states
differing from those present in the lowest mass hadrons, but the
glue holding them in place is still `sticky'. The quarks continue to act as
if still confined within some hadron. This description was successful in say
the Pb+Pb collisions examined in NA49 \cite{NA49}. It remains to be seen
whether at the higher RHIC energies a large fraction of these quarks are free
to roam over large spatial distances, and more importantly perhaps whether
sufficient `free' gluons are present to create the thermodynamic basis for
hadronic material describable as quark-gluon plasma.

Many simulations and/or cascades
\cite{LUCIFERII,wang1,frithjof,werner,genericparton,geiger2,RQMD,URQMD,ARC1}
have been constructed for relativistic heavy ion collisions. Some of these
are purely partonic cascades, some are hybrids of hadronic and partonic
cascading.  LUCIFER \cite{LUCIFERII} is a hadronic cascade run
sequentially through two stages. In the initial rapid phase I, at high
energy, no energy loss is permitted for soft processes; however the complete
collision histories are recorded.  The time duration of phase I, $t_{AB}$, is
essentially that which would be taken by the two colliding nuclei to pass
freely through each other. Hard or partonic processes for which $p_t \ge
t_{AB}^{-1}$ could be introduced in this mode and consequent energy loss
allowed for.

The second stage, phase II, is a conventional hadronic cascade at greatly
reduced energy, similar to that applicable at the AGS and for which soft
energy loss is allowed and chronicled. This second cascade begins only after
a meson formation time, $\tau_f$, has passed. Using the entire space-time and
energy-momentum history of phase I, a reinitialisation is performed using an
elementary hadron-hadron model  fixed by data \cite{LUCIFERII,UA(5),Eisenberg}
as a strict guide. Nucleons travel along  light-cones in phase I,
but the number and type of collisions they suffer are instrumental in
generating the produced mesons which take part in phase II.
Participants in the second phase are treated as generic mesons, thought of as
of $q\bar q$ states with masses centered near $700$ MeV and in the range
$0.3-1.0$ GeV, and generic baryons consisting of $qqq$ excited states also
with rather light masses, $0.94-2.0$ GeV \cite{LUCIFERII}. This same spectrum
of hadrons is of course used to describe the known elementary baryon-baryon
and meson-baryon collisions and the parameters of the model thereby
determined. Ultimately, the cascade is exploited to derive predictions at the
higher energy solely from knowledge of two body interactions and from a
general structure which worked well at the lower $\sqrt{s}\sim 20 \,A$ GeV SPS
level.

In phase II of the ion-ion interaction, the generic resonances decay into
stable hadrons as well as colliding with each other. The low mass
of the generic resonances guarantees that the transverse momentum acquired in
any chain of interactions or decays will be relatively small, and hence one
is modeling only soft processes. A deeper analysis might add parton
production in phase I and cascading perturbatively. Also, and crucially, the
sequential decay of the interacting generic hadrons into several mesons and
baryons severely restricts the particle multiplicities and thus the amount of
cascading during early stages of phase II.  Previously included in our
modeling \cite{LUCIFERI,LUCIFERII} was a suggestion by Gottfried \cite{Gottfried}
that the particles produced in elementary two-body collisions not exist for 
the purpose of secondary interaction until they were sufficiently separated. 
Implementing such a constraint effectively limits the density of
interacting generic hadrons in stage II to non-overlapping configurations. A
very simple but accurate representation of this procedure results from just
constraining the multiplicity at the end of phase I by this criterion, and in
fact the calibration at the NA49 energy $\sqrt{s}=17.2 \,A$ GeV 
then sets the constraint at all energies.

We refer readers to the above mentioned references for more details of the
simulation, the major physical assumptions and measured elementary
hadron-hadron inputs, and the availability of the code. The most important
inputs from involve the total nucleon-nucleon and meson-nucleon
cross-sections and of course the division into elastic, single
diffractive(SD) and non-diffractive production(NSD). The multi-prong UA(5)
\cite{UA(5)} data leading to multiplicity distributions for meson productions
in the latter two categories are crucial.

A concomitant problem in the search for quark-gluon `plasma' is to
distinguish between such a state and simple medium dependence in a hadronic
gas. We constrain the hadronic cascade by imposing no explicit collective
effect of the inter-nuclear environment: however, one could still possibly
ascribe any departure between cascade predictions and measurements to the A
dependences of both particle properties and inter-particle forces on the
conditions obtaining during the nuclear collision. For example, the
apparently anomalous dilepton spectrum at the SPS \cite{spsdilepton} is
frequently attributed to medium dependent shifts in the masses of certain
vector meson resonances \cite{dileptontheory}. Clearly also, particle-particle
cross-sections might be influenced by the presence of a background medium. We
have proceeded however, without introducing any medium dependence whatever.

At RHIC energies the time duration of phase I, $t_{AB} \sim
d_{AB}/\gamma \sim d_{AB}/100$ with $d_{AB}$ being the combined size of
the colliding nuclei, is an order of magnitude shorter than at the
SPS. Moreover, phase II of the cascade at RHIC energies is a more serious
matter. It occurs at higher energies, creates relatively more mesons and
lasts for a longer time. At the SPS \cite{LUCIFERII} we determined the meson
formation time, $\tau_f$, from collisions of light ion systems, e.g S+S,
and we employed this same time, $\tau_f$, in the massive Pb+Pb
system. Inherent in this procedure is the assumed insensitivity of $\tau_f$
to mass number, collision energy, etc. This assumption, equivalent to the one
that hadron properties are independent of the nuclear medium, suggests that
we must use the same $\tau_f$ at RHIC energies, i.~e.  $\tau_f\sim 0.6-0.8$ fm/c.
It would be safer to recalibrate this sensitive parameter, essentially the
only one in our modeling not determined from two body data, with similar
measurements on the light nuclear systems at RHIC. The totality of mesons,
particle and energy densities produced in the cascade are to an appreciable
extent controlled by $\tau_f$, for obvious reasons. For the moment and to
avoid the introduction of any other parameters, we employ the same $\tau_f$
at all energies.

\begin{figure}
\vbox{\hbox to\hsize{\hfil
\epsfxsize=6.1truein\epsffile[27 58 584 494]{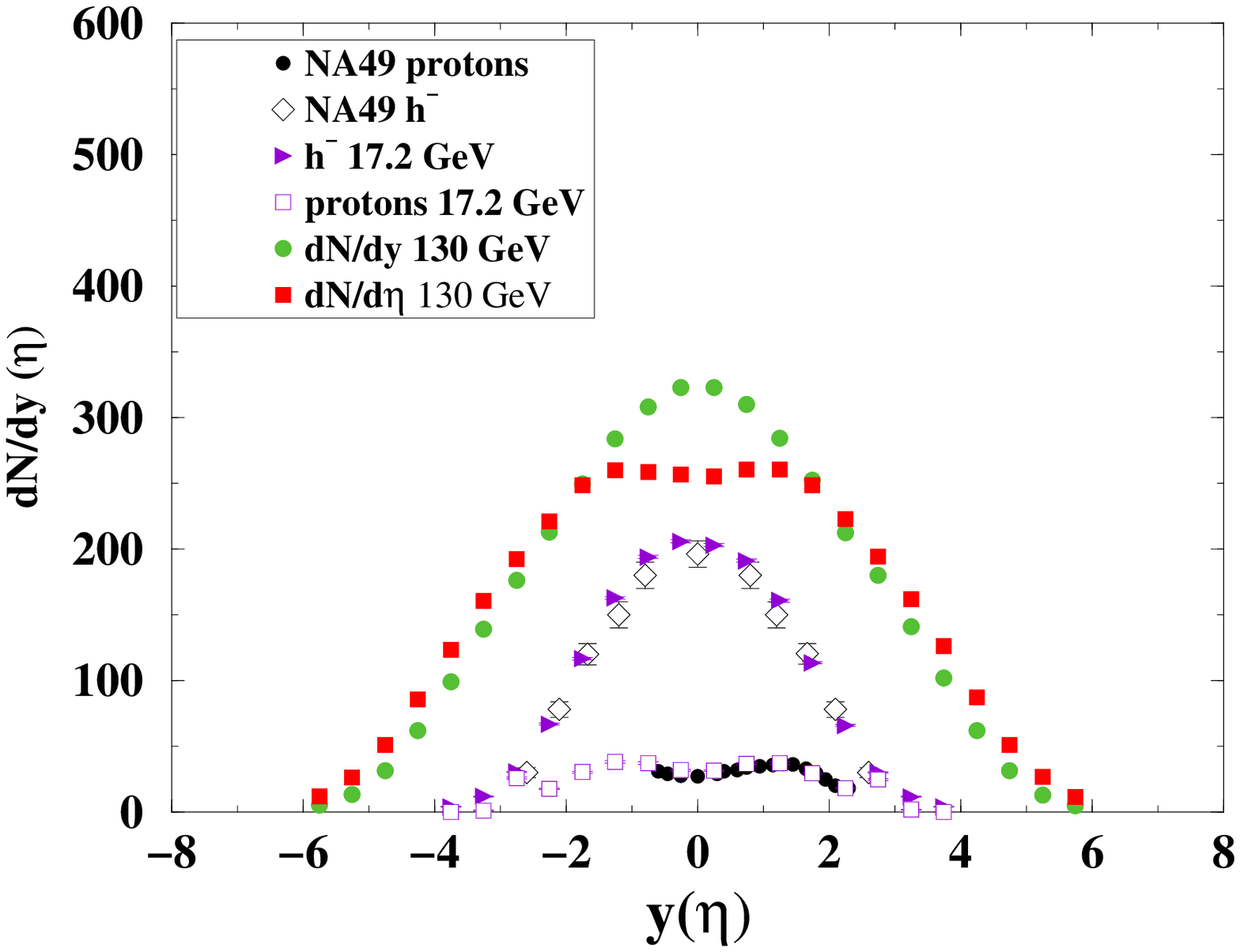}
\hfil}}
\caption[]{Comparison between normalised LUCIFER and NA49 $h^-$ rapidity
 spectra for and protons from Pb+Pb at $158 \,A$ GeV per nucleon (Lab). Also 
 shown are rapidity and pseudo-rapidity distributions for $\pi^-$  
 at $\sqrt{s}=130 \,A$ GeV. The latter should be increased by $\sim 4-5\%$ to 
 include $K^-$ and are not corrected for a possible low $p_t$ cut.}

\label{fig:one}
\end{figure}

To facilitate comparison with the computations at $\sqrt{s}=56-200 \,A$ GeV, we
present here LUCIFER results \cite{LUCIFERII} for Pb+Pb at $E(lab)= 158 \,A$
GeV. These appear in Figure \ref{fig:one} and are there compared to recent
NA49 data \cite{NA49b}. As we described above the code was re-adjusted in
this figure to give near the latest NA49 $\frac{dN}{dy}\vert_{y=0}$ for
negatively charged hadrons, $\pi^-$'s for the most part.

In earlier work \cite{LUCIFERII}, we studied the relativistic invariance of
the model, and demonstrated that for a worst case scenario, i.~e. a zero
impact parameter Au+Au collision at $200 \,A$ GeV, frame dependence in the
cascade, produced by the action at a distance assumptions inherent in the
theory, and as measured by the variation in
$\frac{dN}{dy}_{\pi-}\vert_{y=0}$, was $\le 10\%$, and virtually nonexistent
at SPS energies. Calculations in the present work are performed in the equal
velocity frame for which the errors are undoubtedly less.

We now exhibit typical meson production expected at RHIC in a purely hadronic
simulation. Configurations of the greatest interest involve the most massive
ions in the most central collisions. It is here that one might hope to see
the greatest measured deviations from our simplified purely hadronic, medium
independent picture. The centrality of a collision cannot of course, be
defined in a purely theoretical context; one should account for the complete
experimental set up. However, for simplicity, we specify centrality here by
geometry and initially select $b\le 4$ fm so as to approximately reproduce
the $6\%$ cut specified by PHOBOS \cite{PHOBOS}. Variations in production
levels with impact parameter are not too severe but some error attaches to
the precise definition of centrality. We present results for the two energies
$\sqrt{s}=56,130 \,A$ GeV reported by the PHOBOS Collaboration \cite{PHOBOS},
as well as for the higher RHIC design energy of $\sqrt{s}=200 \,A$ GeV.

The simulation results obtained at SPS energies derived mainly from the above
mentioned inputs: the two body energetics and the totality of nucleon-nucleon
interactions in the course of an event.  Although the time for phase I is
considerably compressed at the higher RHIC energy, we expect much the same
characteristics determine production levels as at the SPS. 

\begin{figure}
\vbox{\hbox to\hsize{\hfil
\epsfxsize=6.4truein\epsffile[27 58 584 494]{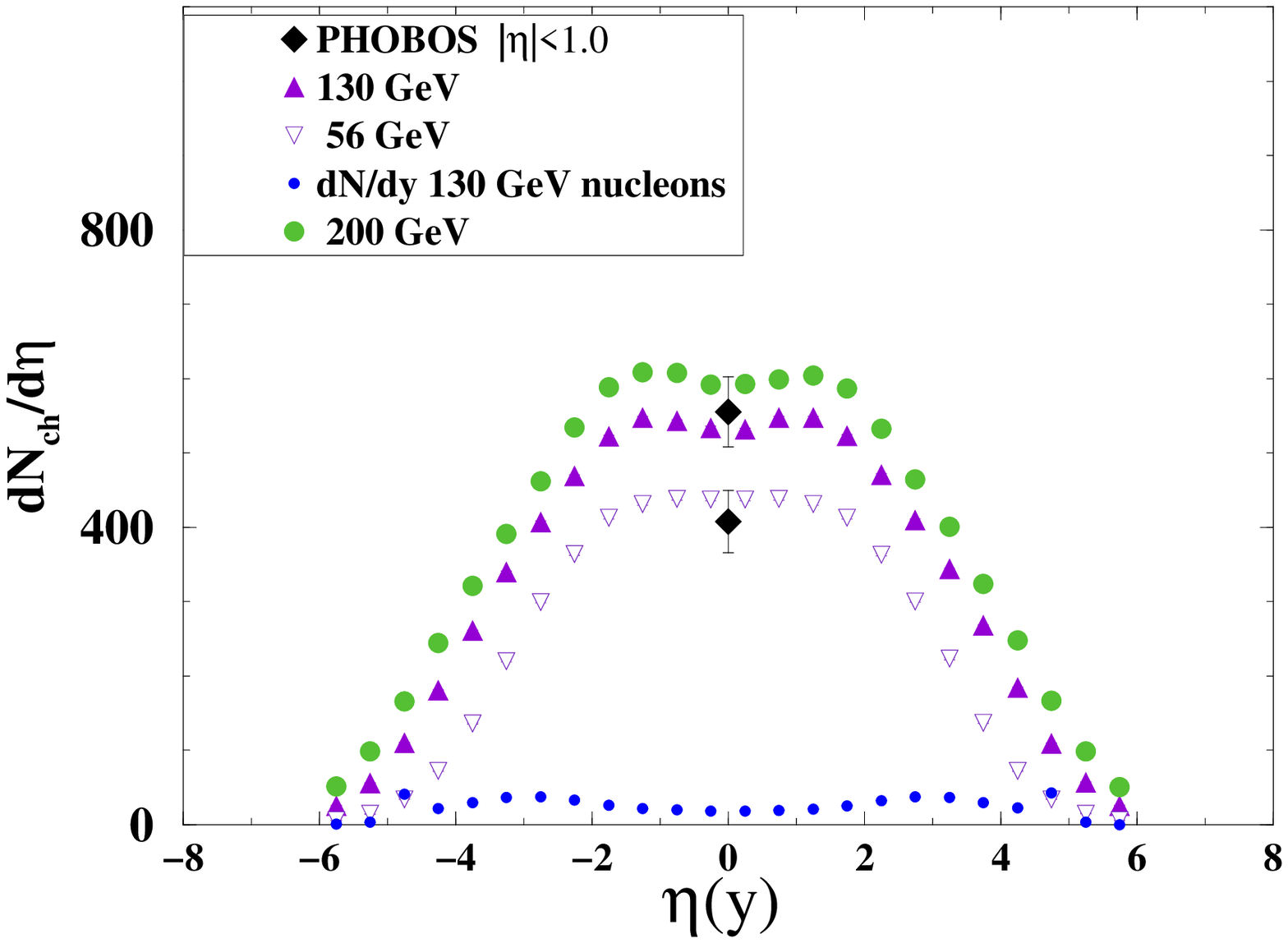}
\hfil}}
\caption[]{Charged Mesons for Au+Au at RHIC energies of $\sqrt{s}=56,130 A$
GeV. Comparison with PHOBOS pseudorapidity averaged density measurements
over the central two units of $\eta$. The LUCIFER spectrum  for
$\sqrt{s}=200 \,A$ GeV is also shown. Small renormalisations can be expected 
for all results from a centrality definition more consistent with individual
experimental setups. The total mesonic production at  $\sqrt{s}=130 \,A$ GeV
in these simulations is some 6500 particles compared to near 2600 at  
$\sqrt{s}=17.2 \,A$ GeV. The nucleon spectrum in this figure is for rapidity 
$y$.}
\label{fig:two}
\end{figure}

The results of the LUCIFER simulations for $\sqrt{s}=56,130 \,A$ GeV are
displayed in Figure \ref{fig:two}, where they are compared to the
corresponding PHOBOS measurements \cite{PHOBOS}. One notes that the PHOBOS
points represent an average over two central units of $\eta$.  The minimum
conclusion to be drawn from the cumulative evidence of Figure \ref{fig:one}
and Figure \ref{fig:two} is surely that LUCIFER provides a satisfactory
explanation of the PHOBOS central rapidity charged meson density
determinations, consistent with the previous normalization of the code to
NA49 data. Additional information contained in Figure \ref{fig:two} is the
predicted energy dependence for the later full energy runs as well as the
shape of $\frac{dN}{d\eta}$ for the complete pseudo-rapidity range. The
precise shape of the spectrum for central $\eta$ yields some information on
the degree of meson cascading. A narrower, less scalloped form suggests a
higher degree of meson re-interaction after formation and/or of baryon-baryon
interaction before. There is perhaps some indication that the theoretical
energy dependence is too muted between $56 A$ and $130 \,A$ GeV, a point to
watch in the as yet unreported results from the other RHIC detectors and as
improvements surely are achieved with ongoing data taking. Similar results of
course can be calculated for the  rapidity ($y$) spectra of each meson species
and for the baryons.

One interesting aspect of the calculations relates to the numbers of
final, observed, mesons resulting with and without phase
II. With the second stage rescattering turned off, all of the final hadrons
are produced from decays of generic resonances that were produced in phase
I. It is on the generic hadrons present after phase I that an effective
multiplicity constraint is placed by normalizing to the SPS data.  This
initial multiplicity directly determines the important early particle and
transverse energy densities. Phase II begins only after a pause, dependent on
$\tau_f$ and the relativistic factors $\gamma$ for the secondary
mesons. Particles produced in phase II begin to materialise only when
the interaction region has increased considerably in size.  The combined
multiplicity increase from phases I+II over phase I with decays is
about a factor $2.25$ at $\sqrt{s}=130 \,A$ GeV. This reasoning suggests that
it is dangerous to tie the final measured $\frac{dN}{d\eta}$, in say PHOBOS,
to an initially achieved $E_T$ density and on this basis to draw the
inference that plasma was formed. Thus the calculated increase in 
particle multiplicity from the SPS to RHIC, $\sim 2.5$, is no sure indicator 
plasma formation is more likely at the higher energy.  
Indeed $\frac{dN}{d\eta}$, which is a better indicator of
central densities during collisions, rises by less than a factor of
$1.4$. One caveat should be placed on the use of $\frac{dN}{d\eta}$ rather
than $\frac{dN}{dy}$.  An examination of Figure 1 shows the calculated ratio
of the central rapidity densities grows by close to a factor of 1.6 from
SPS energy to $\sqrt{s}=130 \,A$ GeV at RHIC, perhaps a more hopeful
circumstance.

The relatively low value of meson density found by PHOBOS is in itself
interpretable as a lack of unusual medium dependence. The increase in entropy
expected from the sudden release of additional parton degrees of freedom ought
to show up as a sharp increase of central $\frac{dN}{d\eta}$ for mesons. Of
course such an increase might yet be present in the neutral mesons, and
mitigating effects like shadowing must be accounted for, but the PHOBOS
$\frac{dN}{d\eta}\vert_{\eta=0}$ must still be considered not unusually high.
Surprises may still arise in the examination of more exclusive observables, 
for example in the very high $p_t$ distributions.

One can now surmise that the anticipated QCD matter behaviour will be, at 
least, harder to detect and must be sought in rarer events. This
conclusion is strengthened by viewing the hadronic cascade as a bridge
between SPS and RHIC energies, with the $\sqrt{s}=17 \,A$ GeV data calibrating
the simulation. In this way the effects of artificialities necessarily
present in this or any theoretical treatments of this complex problem are
lessened. The PHOBOS data was collected for a centrality cut of $6\%$,
reproduced theoretically by selecting $b\leq 4$ fm.  Perhaps one must then
proceed to an order of magnitude higher centrality, e.~g. $\leq 1\%$, or
better still to searching for large multiplicity fluctuations in
order to unearth unusual behaviour. A further observation to be drawn from
our simulations, which will be presented in more detail elsewhere, is that
the hunt for plasma signatures in charmonium suppression is likely to become
increasingly difficult and the quarry to become more elusive at RHIC. The
reason is already evident in present calculations, although the $J/\psi$
survival probability has not yet been estimated. The much larger number of
mesons created in phase II of the LUCIFER simulation at higher energy will
increase the suppression of $J/\psi$. The effective number of `comovers'
has been increased, and the survival after purely hadronic interaction will
be even less.

We have tried adjusting the  dynamics of the simulation  to test the
stability of the extension to higher energy: the inputs and the sharing of
energy among generic resonances. Very little matters aside from  the single
overall normalization of produced particles at the SPS, with small changes
in the latter leading to commensurate effects at RHIC. The broad features of the
free  multiplicity distributions, the energy lost per collision, and thus
deposited in the ion-ion system, together with energy and momentum
conservation seem to be the controlling elements.

Finally, since one could view the LUCIFER cascade as equivalent to a
quark-gluon cascade in which the explicit role of color is neglected, it is
not surprising to find the predictions of apparently widely different
theoretical approaches to be alike \cite{wang1}.

This manuscript has been authored under the US DOE grant
NO. DE-AC02-98CH10886. One of the authors (SHK) is also grateful to the
Alexander von Humboldt Foundation, Bonn, Germany and Hans Weidenmuller,
Max-Planck Institute for Nuclear Physics, Heidelberg for continued support and
hospitality.

\clearpage

\clearpage


\begin{references}

\bibitem{jets}
UA1 Collaboration, C.~Albajar et al., {\it Nucl.~Phys.} {\bf B335}, 261
(1990); G.~Bocquet et al., {\it Phys. Lett.} {\bf B366}, 434, 1996; J.~Ranft,
{\it  Phys.~Rev.} {\bf D51},64 (1995)



\bibitem{eskola}
K.~Eskola, {\it Proceedings, RHIC Summer Study'96}, 99-110, BNL, July 8-19, 1996

\bibitem{infiniteplasma} 
P.~Chen et al. {\it hep-lat/0006010}

\bibitem{earlyuniverse}
A.~M.~Polyakov, {\it Phys. Lett.} {\bf B72}, 477, 1978;
K.~Kajante, C.~Montonen, and E.~Pietarinen, {\it Z. Phys.} {\bf C9}, 253, 1981


\bibitem{sn87a}
S.~H.~Kahana, J.~Cooperstein, and E.~Baron, {\it Phys. Lett.} {\bf B196}, 259, 1987


\bibitem{PHOBOS}
B.~Back et al., the PHOBOS Collaboration,  {\it hep-ex/0007036}

\bibitem{LUCIFERI}
D.~E.~Kahana, {\it Proceedings, RHIC Summer Study'96}, 175-192, BNL, July
8-19, 1996 

\bibitem{LUCIFERII}
D.~E.~Kahana and S.~H.~Kahana, {\it  Phys.~Rev.} {\bf C58}, 3574 (1998);
{\it  Phys.~Rev.} {\bf C59}, 1651 (1999)

\bibitem{NA49}
T.~Wienold and the NA49 Collaboration; In Proceedings of Quark Matter '96, {\it
Nucl. Phys.} {\bf A610}, 76c-87c, 1996; 
P.~G.~Jones and the NA49 Collaboration; In Proceedings of Quark Matter '96, {\it
Nucl. Phys.} {\bf A610}, 76c-87c, 1996;

\bibitem{wang1}
X.~-N.~Wang and M.~Gyulassy, {\it  Phys.~Rev.} {\bf D44}, 3501 (1991); {nucl-th/000814}


\bibitem{NA35} J.~Baechler for the NA35 Collaboration, {\it Phys.~Rev.~Lett.}
{\bf A461}, 72 (1994): S.~Margetis for the NA35 Collaboration, {\it Snowbird
1994, Proceedings, Advances in Nuclear Duynamics}, 128-135, 1994

\bibitem{NA49b}
H.~Appleshauser et al. {\it Phys. Rev. Lett.} {\bf 82}, 2471, 1999.



\bibitem{frithjof}
B.~Andersson, G.~Gustafson, G.~Ingleman, and T.~Sjostrand, {\it Phys.~Rep.} 
{\bf 97}, 31 (1983); B.~Andersson, G.~Gustafson, and B.~Nilsson-Almqvist,
{\it Nucl.~Phys.} {\bf B281}, 289 (1987) 

\bibitem{werner}
J.~Ranft and S.~Ritter, {\it Z.~Phys.} {\bf C27}, 413 (1985); J.~Ranft {\it 
Nucl.~Phys.} {\bf A498}, 111c (1989); A.~Capella and J.~Tran Van, {\it Phys.~
Lett.} {\bf 93B}, 146 (1980) and {\it Nucl.~Phys.} {\bf A461}, 501c (1987):
K.~Werner, {\it Z.~Phys.~C} {\bf 42}, 85 (1989) 



\bibitem{genericparton}
D.~Boal, {\it Proceedings of the RHIC Workshop I}, (1985) and {\it
Phys.~Rev.} {\bf C33}, 2206 (1986); K.~J.~Eskola, K.~
Kajantie and J.~Lindfors, {\it Nucl.~Phys.} {\bf B323}, 37 (1989);

\bibitem{geiger2}
K.~Geiger and B.~Mueller {\it Nucl.~Phys.} {\bf B369}, 600 (1992); K.~
Geiger {\it Phys.~Rev.} {\bf D46}, 4965, and 4986 (1992).
K.~Geiger, {\it Proceedings of Quark Matter'83}, {\it Nucl.~Phys.~ } {\bf
A418}, 257c (1984); K.~Geiger, {\it Phys.~Rev.} {\bf D51}, 2345 (1995)

\bibitem{RQMD}
H.~Stoecker and W.~Greiner, {\it Phys. Rep.} {\bf 137}, 277 (1986);
R.~Mattiello, A.~Jahns, H.~Sorge and W.~Greiner, {\it Phys. Rev, Lett.} {\bf
74} 2180, (1995)

\bibitem{URQMD}
H.~Stoecker, {\it Proceedings, RHIC Summer Study'96}, 211-220, BNL, July
8-19, 1996 

\bibitem{ARC1} Y.~Pang, T.~J.~Schlagel, and S.~H.~Kahana, {\it et al,
Phys.~Rev.~Lett.} {\bf 68}, 2743 (1992); S.~H.~Kahana, D.~E.~Kahana, Y.~Pang,
and T.~J.~Schlagel, {\it Annual Reviews Of Nuclear and Particle Science},
{\bf 46} (1996), (ed C.~Quigg)

\bibitem{UA(5)} G.~Ekspong for the UA5 Collaboration, {\it Nucl.~Phys.} {\bf
A461}, 145c (1987); G.~J.~Alner for the UA5 Collaboration, {\it
Nucl.~Phys.} {\bf B291}, 445 (1987)

\bibitem{Eisenberg}
Y.~Eisenberg {\it et al Nucl.~Phys.} {\bf B154}, 239 (1979)

                                                                          
\bibitem{Gottfried}
K.~Gottfried {\it Phys.~Rev.~Lett.} {\bf 32}, 957 (1974); and {\it Acta.~Phys.
~Pol.} {\bf B3}, 769 (1972)


\bibitem{spsdilepton}
A.~Drees for the CERES/NA45 Collaboration {\it Nucl.~Phys.} {\bf A630}, 449c (1998)

\bibitem{dileptontheory}
C.~M.~Ko, G.~Q.~Li, G.~E.~Brown and H.~Sorge {\it Nucl.~Phys.} {\bf A610}, 342c (1996)

\end{references}
\end{document}